# Speech Command Recognition Using LogNNet Reservoir Computing for Embedded Systems

Yuriy Izotov and Andrei Velichko

Institute of Physics and Technology, Petrozavodsk State University, 33 Lenin St., 185910 Petrozavodsk, Russia
**Corresponding author**
Yuriy Izotov, Institute of Physics and Technology, Petrozavodsk State University, 33 Lenin St., 185910 Petrozavodsk, Russia
E-mail: izotov93.ya@gmail.com
**ORCID iD**
Yuriy Izotov: 0000-0002-4217-7969
Andrei Velichko: 0000-0002-9341-1831

**Abstract.** This paper presents a low-resource speech-command recognizer combining energy-based voice activity detection (VAD), an optimized Mel-Frequency Cepstral Coefficients (MFCC) pipeline, and the LogNNet reservoir-computing classifier. Using four commands from the Speech Commands dataset downsampled to 8 kHz, we evaluate four MFCC aggregation schemes and find that adaptive binning (64-dimensional feature vector) offers the best accuracy-to-compactness trade-off. The LogNNet classifier with architecture 64:33:9:4 reaches 92.04% accuracy under speaker-independent evaluation, while requiring significantly fewer parameters than conventional deep learning models. Hardware implementation on Arduino Nano 33 IoT (ARM Cortex-M0+, 48 MHz, 32 KB RAM) validates the practical feasibility, achieving ~90% real-time recognition accuracy while consuming only 18 KB RAM (55% utilization). The complete pipeline (VAD → MFCC → LogNNet) thus enables reliable on-device speech-command recognition under strict memory and compute limits, making it suitable for battery-powered IoT nodes, wireless sensor networks, and hands-free control interfaces.

## 1 Introduction

Voice command recognition on microcontroller-based systems is rapidly becoming an essential technology, enabling intuitive, hands-free control for diverse applications ranging from smart home automation and robotics to industrial and assistive devices. The primary challenge in this domain lies in implementing sophisticated neural network models within the stringent computational constraints typical of embedded platforms.

As a result, significant research efforts have been directed toward developing architectures and feature extraction methods that balance high recognition accuracy with minimal hardware requirements.

Recent literature highlights several neural network architectures optimized for deployment on resource-constrained microcontrollers. Among these, convolutional neural networks (CNNs) [1] and depthwise separable CNNs (DSCNNs) [2] have proven particularly effective, achieving accuracies exceeding 90% with low power consumption and fast inference times [3, 4]. DSCNNs specifically provide computational efficiency, making them suitable for bare-metal microcontroller implementations, often reaching accuracy around 91% even with minimal computational resources [2]. Additionally, recurrent neural networks (RNNs) [5], including hybrid architectures such as convolutional recurrent neural networks (CRNNs) [6], are used to model temporal dependencies in speech signals, although these networks typically demand more computational resources and require careful optimization [7].

An essential component influencing recognition performance on microcontrollers is the feature extraction stage, where Mel-Frequency Cepstral Coefficients (MFCCs) remain the most widely adopted approach due to their effectiveness in capturing crucial speech characteristics [8, 9]. Various aggregation methods have been explored to enhance the temporal dynamics representation of MFCC features. Static feature aggregation provides reasonable accuracy (86%), but advanced approaches such as temporal window statistics and adaptive binning methods demonstrate significantly higher performance (up to 95%) by effectively preserving time-domain information within a compact feature representation [10].

Given the limited processing capabilities of microcontrollers, researchers have adopted numerous optimization techniques to reduce computational and memory requirements. Model quantization (e.g., reducing precision to 8- or 16-bit integers) and pruning have become standard practices, significantly lowering memory usage without substantial loss of accuracy [11, 12]. Frameworks like TensorFlow Lite for Microcontrollers have been widely employed to facilitate such optimizations, promoting ease of deployment and practical implementation on embedded hardware.

This paper introduces a voice command recognition system combining an efficient voice activity detection algorithm, optimized MFCC feature extraction methods, and the LogNNet neural network architecture based on reservoir computing [13, 14]. The LogNNet architecture was specifically chosen due to its potential for maintaining high accuracy while significantly reducing computational load, thus making it particularly suitable for embedded systems. The primary aim of our study is to achieve high recognition accuracy of basic command sets with minimal computational requirements, ensuring suitability for implementation on embedded platforms.

## 2    Methods

### 2.1    Description of the dataset and data preprocessing.

The Speech Commands dataset [15] was utilized as the primary data source in this research. This dataset comprises approximately 65,000 one-second audio utterances of 30 distinct short words, recorded by thousands of different speakers. From this extensive collection, four specific control commands — 'go', 'stop', 'left', and 'right' — were selected, guided by the research goal of implementing real-time device control systems.

For each selected command, the following data distribution was used: 'go' (3880 samples), 'stop' (3872 samples), 'left' (3801 samples), and 'right' (3778 samples), totaling 15331 samples. The dataset was partitioned using an 80/20 split, resulting in a training set of 12265 samples (80%) and a test set of 3066 samples (20%).

Additionally, the inherent presence of diverse background noises and significant variability in speaker characteristics within the dataset enhances the robustness of the obtained results, thereby ensuring higher resilience to environmental disturbances, noise interference, and anomalous inputs.

The original audio samples within the dataset were recorded at a sampling frequency of 16 kHz. To optimize computational complexity and improve processing efficiency, the signals underwent resampling from 16000 Hz to 8000 Hz. This decision effectively reduced the dimensionality of input data by half while preserving essential spectral characteristics relevant to speech command recognition.

Voice Activity Detection (VAD) based on an energy thresholding technique was applied for segment extraction. The VAD algorithm parameters were empirically optimized through preliminary experiments on a subset of the training data, with the following configuration: analysis window length of 1000 samples (125 ms at 8 kHz), window shift step of 300 samples (37.5 ms at 8 kHz), energy threshold of 0.001 (normalized Mean Squared Energy), acceptable command duration range of 0.1–0.7 seconds, and temporal padding of 0.05 seconds before and after the detected speech segment.

The audio signal was divided into frames of 1000 samples, with each frame shifted forward by 300 samples. Within each frame, the Mean Squared Energy (MSE) was calculated as:

$$E_{MSE} = \frac{1}{N} \sum_{i=1}^{N} x_i^2 \tag{1}$$

where $x_i$ represents the amplitude of each audio sample within the frame, and NNN is the number of samples per frame. This method ensures precise localization of speech activity within the audio signal. Segments with energy exceeding the predefined threshold were identified as active speech segments. The onset of each segment was marked when the energy first surpassed the threshold, and the offset was noted when the energy fell below the threshold. The initial and terminal frames of each segment were tracked accordingly.

Each detected segment was validated against the predefined temporal constraints. In instances where multiple segments met the criteria, the segment with the highest mean energy was selected, calculated as follows:

$$\bar{E} = \frac{1}{M} \sum_{j=1}^{M} E_{MSE_j} \tag{2}$$

where $E_{mse}$ is the MSE for each frame within the segment, and $M$ is the total number of frames comprising the segment. This criterion ensures selection of the most clearly articulated command in scenarios with multiple activations. Additionally, temporal padding of 0.05 seconds was appended before and after the selected segment to preserve transitional features present at the segment boundaries.

The principle of the algorithm is visualized in Figure 1, which shows the original sound waveform, its corresponding energy representation, the energy threshold line, and the time range of the extracted segment of the "go" command.

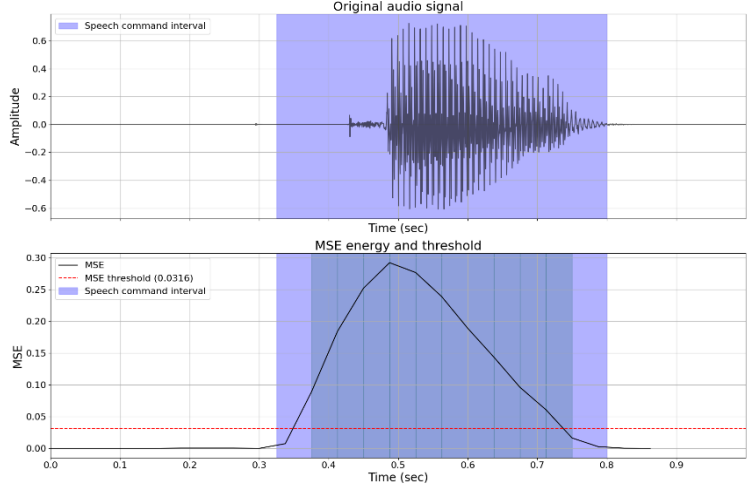

**Fig. 1.** Visualization of VAD algorithm operation: original audio signal (command – «go»), energy profile, threshold level, and detected speech command interval.

Following the extraction of speech segments via VAD, Mel-Frequency Cepstral Coefficients (MFCCs) were computed for each isolated command, providing a compact and effective feature representation for further classification tasks

### 2.2   LogNNet Classifier Principle of Operation

The LogNNet neural network represents a reservoir computing-based architecture that leverages chaotic dynamical systems for efficient feature transformation. The input feature vector $F$ comprises $N$ elements ($F_1, F_2, ..., F_n$), where $N$ ranges from 32 to 128 depending on the aggregation method. The classifier outputs one of four classes corresponding to the audio commands ('go', 'left', 'right', 'stop').

The structural diagram illustrating the LogNNet neural network architecture is depicted in Fig. 2. It consists of a reservoir characterized by a specialized matrix, denoted as $W$. Matrix $W$ is filled row-wise with numbers generated by a chaotic mapping $x_n$. Initially, the feature vector $F$ is transformed into a vector $Y$ of dimensionality $N+1$, introducing an additional coordinate $Y_0=1$ (bias). Each component of this augmented vector is then normalized by dividing by the maximum value of that component observed in the training dataset. Subsequently, the matrix $W$ is multiplied by vector $Y$, yielding an intermediate vector $S$ of dimension $P$. Vector $S$ is further normalized and expanded into vector $Sh$ of dimension $P+1$, incorporating a bias term $Sh[0]=1$. Thus, an initial transformation of the feature vector *F* into a second, (P+1)-dimensional space is achieved.

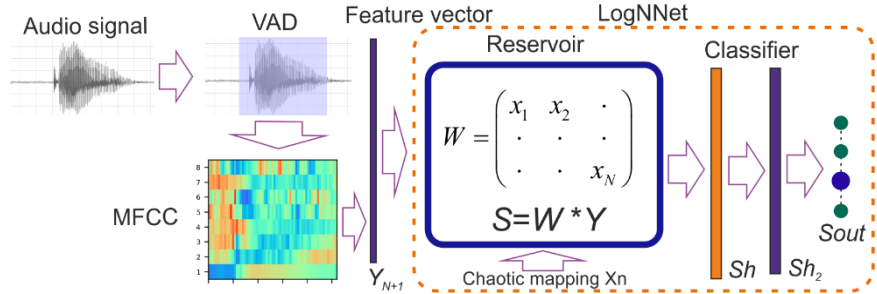

**Fig. 2.** General scheme of the LogNNet neural network.

Finally, the transformed vector *Sh* serves as the input to a two-layer linear classifier. This classifier comprises a hidden layer $Sh_2$ containing *M* neurons and an output layer Sout with 4 neurons, corresponding to the number of classes. The concise notation LogNNet N:P:M:4 is used to denote the parameters defining the structure of the neural network.

## 2.3 Extraction of MFCC features and aggregation of the resulting vector.

The MFCC extraction parameters were specifically optimized for processing short spoken commands within the scope of this study. As noted above, the signals were sampled at 8000 Hz. The FFT length was set to 128 samples, which at the chosen sampling rate corresponds to a temporal window of approximately 16 ms and a frequency resolution of 62.5 Hz. The analysis hop size was 64 samples (8 ms), i.e., 50% overlap between adjacent frames. A bank of 12 mel filters was used, providing adequate spectral detail on the mel scale while preserving computational efficiency.

We computed 9 MFCC coefficients per frame and retained the first 8 (discarding the highest-order coefficient as mostly noise-dominated). This number of coefficients provides a balance between capturing enough spectral detail and maintaining computational efficiency: too few coefficients might miss important acoustic information, while too many would add unnecessary complexity without improving classification accuracy. Also, higher-order (high-frequency) cepstral coefficients predominantly reflect rapidly varying spectral components and often contain noise and discretization artifacts that do not contribute meaningful information for subsequent processing. Fig. 3 shows the MFCC coefficient matrix for all sound commands used in the study. The color scale represents z-scored coefficient amplitudes; panel titles indicate the corresponding command labels.

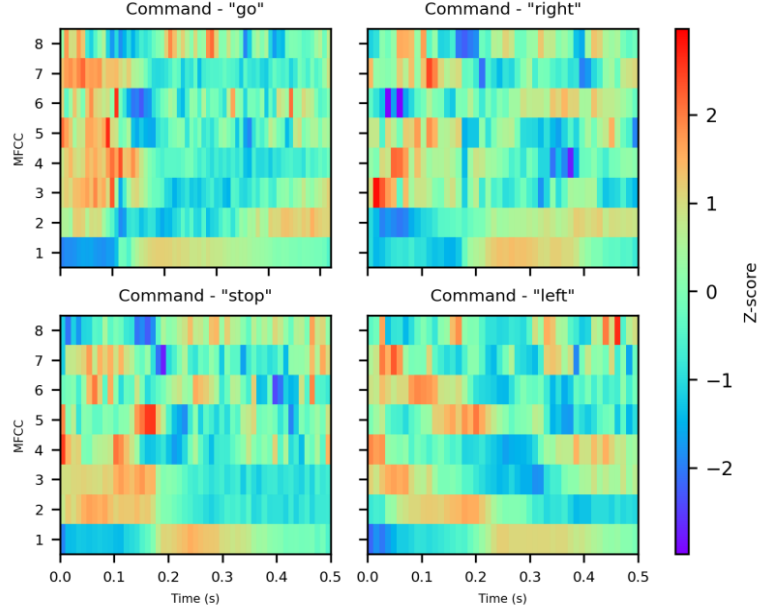

**Fig. 3.** Z-score normalized MFCC coefficient matrices for four example spoken commands.

Before feeding the MFCC feature matrix into a neural-network classifier, the matrix must be converted into a one-dimensional feature vector. In this study, four aggregation methods for performing this conversion were implemented and evaluated.

**Basic statistical features method.**
This method is based on computing basic statistical descriptors for each MFCC coefficient along the time axis. For each *i*-th coefficient *mfcc_i(t)*, where *t = 1, 2, ..., N* (N is the number of time frames), the following statistics are computed:

$$\begin{aligned}
\mu_i &= \frac{1}{N} \sum_{t=1}^{N} mfcc_i(t) \\
\sigma_i &= \sqrt{\frac{1}{N-1} \sum_{t=1}^{N} (mfcc_i(t) - \mu_i)^2} \\
min_i &= \min_{t \in [1,N]} mfcc_i(t) \\
max_i &= \max_{t \in [1,N]} mfcc_i(t)
\end{aligned} \quad (3)$$

This approach effectively compresses the temporal dynamics of each coefficient into a compact representation of four values, resulting in the formation of a feature vector with dimensionality of 32 features (8 coefficients × 4 statistics). The method is characterized by high computational efficiency and ease of implementation on embedded platforms.

**Temporal dynamics method.**
The method is based on analyzing the temporal dynamics of MFCC coefficients by computing first and second derivatives, which enables capturing information about the velocity and acceleration of spectral characteristic changes. For each coefficient *mfcc_i(t)* are calculated:

Basic statistics. The mean $\mu_i$ and the standard deviation $\sigma_i$ are defined as:

$$\mu_i = \frac{1}{N}\sum_{t=1}^{N} mfcc_i(t)$$
$$\sigma_i = \sqrt{\frac{1}{N-1}\sum_{t=1}^{N}(mfcc_i(t)-\mu_i)^2} \quad (4)$$

First derivative (delta coefficients):

$$\Delta mfcc_i(t) = mfcc_i(t+1) - mfcc_i(t), \quad t=1,2,...,N-1$$
$$\mu_{\Delta i} = \frac{1}{N-1}\sum_{t=1}^{N-1}\Delta mfcc_i(t) \quad (5)$$
$$\sigma_{\Delta i} = \sqrt{\frac{1}{N-2}\sum_{t=1}^{N-1}(\Delta mfcc_i(t)-\mu_{\Delta i})^2}$$

Second derivative (delta-delta coefficients):

$$\Delta^2 mfcc_i(t) = \Delta mfcc_i(t+1) - \Delta mfcc_i(t), \quad t=1,2,...,N-2$$
$$\mu_{\Delta^2 i} = \frac{1}{N-2}\sum_{t=1}^{N-2}\Delta^2 mfcc_i(t) \quad (6)$$
$$\sigma_{\Delta^2 i} = \sqrt{\frac{1}{N-3}\sum_{t=1}^{N-2}(\Delta^2 mfcc_i(t)-\mu_{\Delta^2 i})^2}$$

As a result, the resulting vector with a dimension of 48 features (8 coefficients × 6 statistics) is formed as:

$$F = [\mu_1, \sigma_1, \mu_{\Delta 1}, \sigma_{\Delta 1}, \mu_{\Delta^2 1}, \sigma_{\Delta^2 1}, ..., \mu_8, \sigma_8, \mu_{\Delta 8}, \sigma_{\Delta 8}, \mu_{\Delta^2 8}, \sigma_{\Delta^2 8}] \quad (7)$$

This approach provides a more comprehensive representation of temporal changes in the spectral characteristics of the speech signal.

**Windowed statistical method.**
The method implements analysis of local temporal characteristics by partitioning the MFCC matrix into a fixed number of temporal windows followed by computation of statistical descriptors within each window. The MFCC matrix is divided into 4 equal temporal segments. For each window w and each coefficient *i*, the following are computed:

$$\mu_i^{(k)} = \frac{1}{t_{end}^{(k)} - t_{start}^{(k)} + 1} \sum_{t=t_{start}^{(k)}}^{t_{end}^{(k)}} mfcc_i(t)$$

$$\sigma_i^{(k)} = \sqrt{\frac{1}{t_{end}^{(k)} - t_{start}^{(k)}} \sum_{t=t_{start}^{(k)}}^{t_{end}^{(k)}} (mfcc_i(t) - \mu_i^{(k)})^2} \qquad (8)$$

$$min_i^{(k)} = \min_{t \in [t_{start}^{(k)}, t_{end}^{(k)}]} mfcc_i(t)$$

$$max_i^{(k)} = \max_{t \in [t_{start}^{(k)}, t_{end}^{(k)}]} mfcc_i(t)$$

These statistical features capture the temporal dynamics of the spectral characteristics within each segment, providing a comprehensive representation of the signal's time-varying properties. As a result, the resulting vectors with a dimension of 128 features (8 coefficients × 4 windows × 4 statistics) is formed as:

$$F = [\mu_1^{(1)}, \sigma_1^{(1)}, min_1^{(1)}, max_1^{(1)}, \mu_2^{(1)}, \sigma_2^{(1)}, min_2^{(1)}, max_2^{(1)}, ..., \mu_N^{(k)}, \sigma_N^{(k)}, min_N^{(k)}, max_N^{(k)}] \qquad (9)$$

**Adaptive binning method.**
The method is based on uniform partitioning of the temporal axis of each MFCC coefficient into a fixed number of intervals (bins) with computation of the mean value within each interval. For each coefficient *mfcc_i(t)*, the temporal axis is divided into B = 8 equal intervals. For each interval b, the mean value is computed as:

$$bin_i^{(b)} = \frac{1}{t_{end}^{(b)} - t_{start}^{(b)} + 1} \sum_{t=t_{start}^{(b)}}^{t_{end}^{(b)}} mfcc_i(t) \qquad (10)$$

The term "adaptive" reflects the method's ability to automatically adjust the size of temporal intervals to the duration of the input signal, ensuring a fixed number of features regardless of recording length. The resulting vector is calculated as:

$$F = [bin_1^{(1)}, bin_1^{(2)}, ..., bin_1^{(8)}, bin_2^{(1)}, bin_2^{(2)}, ..., bin_2^{(8)}, ..., bin_8^{(1)}, bin_8^{(2)}, ..., bin_8^{(8)}] \qquad (11)$$

The vector contains 64 features (8 coefficients × 8 intervals). This approach provides a compromise between temporal representation detail and computational efficiency

## 2.4 Hardware implementation on the Arduino Nano 33 IoT board.

To experimentally verify the proposed algorithm under limited computational resources, a fully functional voice command recognition system was implemented on the Arduino Nano 33 IoT platform. This platform features a Microchip SAMD21G18 microcontroller with a 32-bit ARM Cortex-M0+ processor core running at 48 MHz clock frequency, equipped with 256 KB of flash memory and 32 KB of RAM. The module additionally includes an ATECC608A cryptoprocessor for hardware security and a u-blox NINA-W102 wireless communication module with Wi-Fi and Bluetooth Low Energy support, enabling device integration into distributed IoT networks.

A MAX9814 microphone module with integrated automatic gain control (AGC) was used as the audio source, providing stable signal levels at varying speech volumes and suppressing abrupt amplitude changes. The microphone output was connected to analog input A0 of the Arduino board, where the signal was digitized by the built-in 12-bit ADC at a sampling rate of 8 kHz. To ensure precise sampling frequency, hardware timer TC3 was configured to generate interrupts with a period of 125 μs. With a 48 MHz clock frequency and prescaler of 64, the compare register value was set to 93.75, providing an interrupt frequency of:

$$f_{int} = \frac{f_{clk}}{DIV \times (CC + 1)} = \frac{48\ MHz}{64 \times 93.75} \approx 7979\ Hz \qquad (12)$$

where $f_{clk}$ = 48 MHz is the system clock frequency, DIV = 64 is the prescaler division factor that reduces the input clock frequency to the timer counter (effectively dividing the 48 MHz system clock by 64 to obtain 750 kHz timer clock), and CC = 93.75 is the compare register value that determines the timer overflow point (the timer counts from 0 to CC and generates an interrupt when reaching this value, then resets to 0).

The continuous audio data stream was stored in a circular buffer of 4000 samples (0.5 seconds), allowing capture of the audio signal preceding command detection while preserving transient characteristics at the beginning and end of speech segments without introducing processing delays. Voice activity detection was implemented using an energy-based threshold method with a finite state machine having four states: SILENCE, MAYBE_SPEECH, SPEECH, and MAYBE_SILENCE. Energy was calculated for each window of 160 samples (20 ms) with a shift of 40 samples (5 ms) according to equation (1), where normalization was performed by dividing by 2048, corresponding to half the dynamic range of the 12-bit ADC.

The adaptive detection threshold was determined as the product of the current noise level estimate and empirically selected multipliers: 6.0 for speech onset and 2.5 for speech offset. These coefficients were chosen to ensure reliable command detection at signal-to-noise ratios of at least 15 dB, typical for indoor environments, while minimizing false detections. The higher threshold for speech onset prevents response to brief impulse noise, while the lower offset threshold allows correct determination of command boundaries with decaying amplitude. Segment duration was limited to the range of 300-700 ms, with 50 ms of silence after sound cessation considered as a command completion indicator.

After speech segment extraction, MFCC feature extraction was performed directly on the microcontroller. Processing was carried out frame by frame using a Hamming window of 128 samples (16 ms at 8 kHz) with 50% overlap between adjacent frames. The Hamming window was applied to reduce spectral leakage. This window function was chosen as it provides better side lobe suppression compared to a rectangular window (-13 dB) with acceptable main lobe broadening.

For each windowed frame, a 128-point FFT was performed followed by amplitude spectrum calculation. The resulting spectrum was transformed to the mel scale using a bank of 12 triangular filters uniformly distributed in the range of 300-3800 Hz. Filter energies were logarithmically transformed followed by discrete cosine transform application for decorrelation, resulting in a vector of 8 MFCC coefficients for each time frame.

Classification was performed using a pre-trained LogNNet neural network with architecture N:P:M:4, where the first layer represents a reservoir with pseudo-random weights generated by a linear congruential generator [16], and subsequent layers form a fully connected perceptron with tanh activation for the reservoir and ReLU for hidden layers. The network weights and normalization parameters were stored in program memory using the PROGMEM directive, significantly reducing RAM usage.

## 3  Results.

### 3.1  Evaluation of classification efficiency.

For objective evaluation of the proposed voice command recognition system, two data splitting strategies were employed. The first was random splitting in an 80/20 ratio, while the second was speaker-independent splitting, where speakers voice from the test set were completely absent from the training set. The second approach is the standard for evaluating speech recognition systems, as it excludes overfitting to specific speaker's vocal characteristics.

Experiments revealed a significant difference between the two approaches. For the adaptive binning method, accuracy with speaker-independent splitting was 92.04%, whereas with random splitting it was 94.64%. The 2.6% difference is explained by the fact that with random splitting, the same speaker's voice can appear in both training and test sets, simplifying the classification task and providing an inflated accuracy estimate.

Table 1 presents a comparison of classification accuracy under different data splitting strategies for all investigated MFCC feature aggregation methods. The basis was the neural network LogNNet with the architecture N:50:40:4, where N is the length of the input vector.

Table 1. Comparison of accuracy metrics for different aggregation methods.

| Aggregation method | Vector dimension (N) | Speaker-Independent | Random Split 80/20 | Overestimation |
|---|---|---|---|---|
| Basic statistical features method | 32 | 0.8249 | 0.8548 | +2.99% |
| Temporal dynamics method | 48 | 0.8576 | 0.8722 | +1.46% |
| Windowed statistical method | 128 | 0.9172 | 0.9380 | +2.08% |
| Adaptive binning method | 64 | 0.9204 | 0.9464 | +2.60 % |

As shown in the table, all methods demonstrate inflated accuracy with random splitting. The overestimation varies from 1.46 % to 2.99 %, confirming the importance of using the speaker-independent approach for realistic system evaluation.

The adaptive binning method showed the best result (92.04 %) with a moderate feature vector dimension (64 elements). This is an important advantage for embedded systems where memory and computational resources are limited. The windowed statistical method provides similar accuracy (91.72 %), but requires twice the memory for storing the feature vector (128 elements).

The basic statistical features method, despite its compactness (32 features), shows accuracy of only 82.49 %, which may be insufficient for practical applications. The temporal dynamics method occupies an intermediate position in both dimensionality (48 features) and accuracy (85.76 %).

Table 2 presents class-wise precision, recall, and F1-scores for the best-performing adaptive binning method under both data splitting strategies.

**Table 2.** Detailed classification metrics for the adaptive binning method.

| Command | Speaker-Independent Split | | | Random Split 80/20 | | |
| --- | --- | --- | --- | --- | --- | --- |
| | Precision | Recall | F1-score | Precision | Recall | F1-score |
| go | 0.8981 | 0.9261 | 0.9119 | 0.9514 | 0.9497 | 0.9505 |
| left | 0.9211 | 0.9211 | 0.9211 | 0.9342 | 0.9436 | 0.9389 |
| right | 0.9194 | 0.9087 | 0.9140 | 0.9428 | 0.9404 | 0.9416 |
| stop | 0.9431 | 0.9251 | 0.9340 | 0.9572 | 0.9519 | 0.9545 |
| Macro avg | 0.9204 | 0.9203 | 0.9202 | 0.9464 | 0.9464 | 0.9464 |

The comparison clearly demonstrates the systematic improvement in all metrics when using random splitting, with F1-scores increasing by 0.025-0.039 across all classes. However, this improvement is misleading as it results from data leakage rather than genuine model performance.

Beyond basic accuracy metrics, several additional indicators were calculated to provide a more comprehensive assessment of classifier performance. The balanced accuracy of 0.9203 for speaker-independent splitting, though lower than the 0.9464 obtained with random splitting, represents the true system capability when encountering unseen speakers. The Matthews Correlation Coefficient (MCC) of 0.8939 for speaker-independent evaluation indicates strong correlation between predicted and actual classes under realistic conditions.

**Table 3.** Summary of extended metrics for all aggregation methods.

| Method | Accuracy | Balanced Accuracy | MCC | Macro Precision | Macro Recall | Macro F1 |
| --- | --- | --- | --- | --- | --- | --- |
| Basic statistical features | 0.8249 | 0.8248 | 0.7665 | 0.8260 | 0.8248 | 0.8250 |
| Temporal dynamics | 0.8576 | 0.8574 | 0.8101 | 0.8580 | 0.8574 | 0.8575 |
| Windowed statistics | 0.9172 | 0.9171 | 0.8896 | 0.9180 | 0.9171 | 0.9172 |
| Adaptive binning | 0.9204 | 0.9203 | 0.8939 | 0.9204 | 0.9203 | 0.9202 |

These comprehensive metrics confirm that the adaptive binning method provides the best overall performance with accuracy of 92.04% and MCC of 0.8939, marginally outperforming the windowed statistics method while using only half the feature vector dimensionality (64 vs 128 features). The high balanced accuracy (92.03%) demonstrates that this performance is consistent across all command classes, making the system reliable for practical deployment.

### 3.2   Feature Importance Analysis and Selection.

To ensure model interpretability and optimize the feature set for embedded systems, a Permutation Feature Importance (PFI) analysis was conducted. This method quantitatively evaluates each feature's contribution to the final classification accuracy by measuring performance degradation when randomly permuting specific feature values on the test set. The PFI analysis was performed for all four MFCC feature aggregation methods using the LogNNet N:50:40:4 architecture, where N corresponds to the input vector dimensionality. To ensure statistical reliability, 3-fold cross-validation with speaker-stratified splitting was employed. At each cross-validation iteration, 10 independent permutations were performed for each feature, followed by computation of mean accuracy drop.

The analysis results for the adaptive binning method (see Fig. 4) demonstrate a pronounced non-uniform distribution of feature importance with clear dominance of the second MFCC coefficient. The fourth bin of this coefficient causes the maximum accuracy drop of 0.0407, representing 20.6% of the cumulative importance of the top-10 features and indicating the critical role of the command's central portion in the recognition process. The seventh bin of the same coefficient ranks second in importance with a drop of 0.0214, confirming the informativeness of the utterance's final phase. Notably, 70% of the most important features are concentrated in bins three through seven, while initial bins are virtually absent among critically important features, which aligns with the characteristics of short command utterances where primary phonetic information is localized in central and final portions.

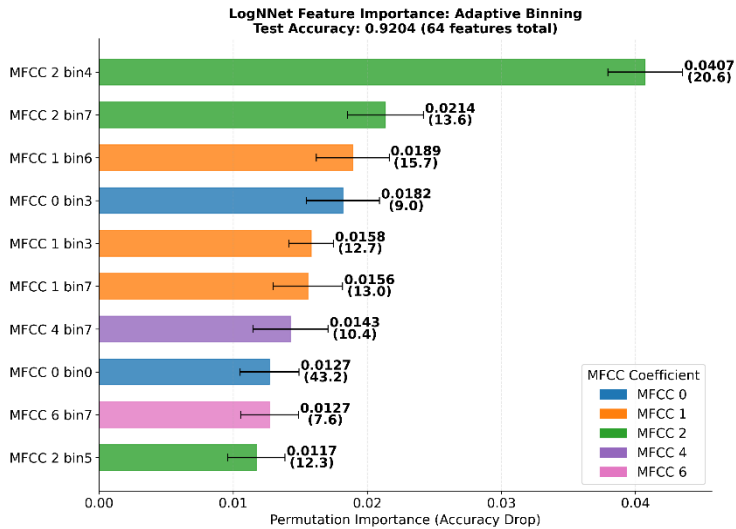

**Fig. 4.** Permutation Feature Importance analysis for the adaptive binning method.

Based on the PFI analysis, a progressive feature reduction procedure was conducted to optimize the model for embedded system constraints. For the adaptive binning method, which demonstrated the best accuracy-to-compactness ratio, an iterative feature selection algorithm was implemented. At each iteration, features with the lowest importance were removed, after which the model was retrained and evaluated on the test set.

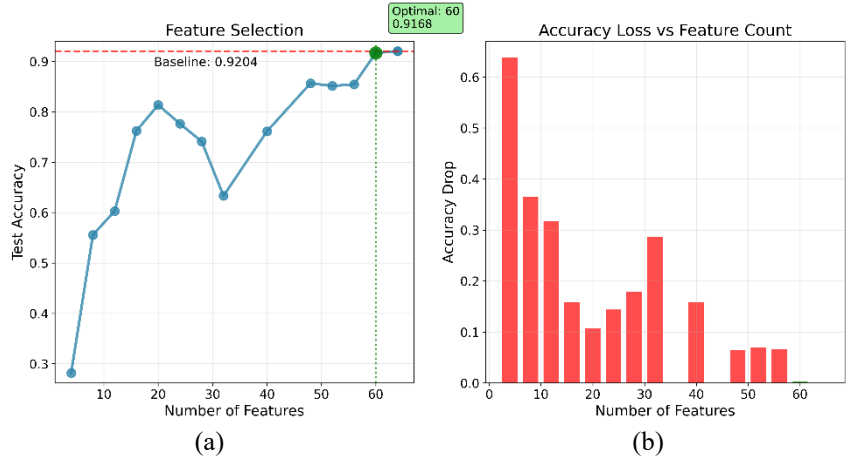

(a) (b)

**Fig. 5.** Classification accuracy dependence on the number of features used for the adaptive binning method. (a) – accuracy curve with marked optimal configuration. (b) – absolute accuracy drops relative to baseline level.

Analysis of the feature reduction curve (see Fig. 5) revealed a complex nonlinear relationship between accuracy and dimensionality. Using the minimal set of four features yields only 28.2% accuracy, indicating critical information insufficiency for distinguishing four command classes. Of particular interest is the unstable zone in the 16-32 feature range, where pronounced accuracy fluctuations occur with a local maximum of 81.7% at 20 features and an unexpected drop to 63.9% at 32 features. Such behavior indicates the suboptimality of the greedy selection algorithm, which fails to account for synergistic effects between features and may include mutually correlated characteristics that degrade the model's generalization capability.

Performance stabilization occurs only after 40 features, where accuracy reaches 76.21% and then monotonically increases to 86.03% at 52 features. Achieving accuracy within 2% of the baseline level is possible only when using 60 or more features. At this configuration, accuracy reaches 91.68%, merely 0.36% below baseline, while the complete set of 64 features ensures maximum accuracy of 92.04%.

The obtained results demonstrate limitations in feature space reduction possibilities for the adaptive binning method. Contrary to initial expectations of achieving an effective configuration with half the features, experiments showed that removing more than four least important characteristics leads to statistically significant performance degradation. This can be explained by adaptive binning already representing a compact representation of MFCC coefficient temporal dynamics, where each feature carries substantial information about the command's spectro-temporal structure.

### 3.3  Optimization of LogNNet Architecture Parameters.

Additionally, an experiment was conducted to investigate the dependence of classification performance on the size of the layers in the LogNNet neural network with architecture N:P:M:4. In this experiment, the parameters P — the number of rows in the reservoir — and M — the number of neurons in the hidden layer — were varied in the range from 1 to 50. The results of this study are presented in Fig. 6.

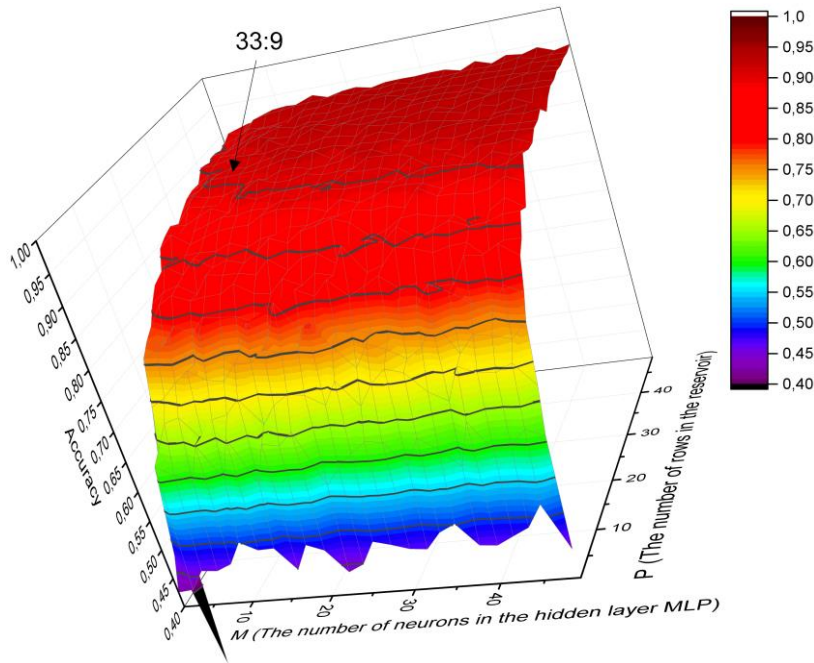

**Fig. 6.** A three-dimensional plot of the classification accuracy of the LogNNet classifier as a function of the reservoir size (P) and the number of neurons in the hidden layer (M) is shown.

The experiment revealed a proportional increase in classification accuracy with increasing P and M values. A local optimum was identified where the LogNNet network with architecture 64:33:9:4 achieved an accuracy of approximately 90 %. Beyond this point, further increases in P and M resulted only in higher computational load, while the accuracy improvement was marginal (approximately 2 % when increasing to 64:50:40:4). This optimization directly influenced the selection of the neural network architecture for the microcontroller.

### 3.4 Embedded Platform Implementation.

For practical verification of the algorithm, a complete implementation was performed on the Arduino Nano 33 IoT microcontroller with ARM Cortex-M0+ processor (48 MHz, 32 KB RAM). Based on the optimization results presented in Figure 6, the LogN-Net architecture 64:33:9:4 was selected for the embedded implementation, providing an optimal balance between classification accuracy and computational cost.

The implementation follows a three-stage processing pipeline: Voice Activity Detection (VAD), MFCC feature extraction, and LogNNet classification, Practical tests showed that the system reliably recognizes four target commands ('go', 'stop', 'left', 'right') under moderate background noise conditions. Recognition accuracy on the embedded platform reached ~90%, which is slightly lower than the results obtained in PC simulation (~92%). This is explained by the use of a simplified neural network architecture and limited floating-point computation precision on a microcontroller.

— Stage 1: Voice Activity Detection (VAD).

The VAD module continuously monitors the incoming audio stream stored in a 4000-sample circular buffer (8 KB). Energy calculation is performed every 40 samples (5 ms) using a sliding window of 160 samples. The finite state machine transitions between four states (SILENCE, MAYBE_SPEECH, SPEECH, MAYBE_SILENCE) based on adaptive energy thresholds. Upon detecting a valid speech segment (300-700 ms duration), the module triggers the MFCC extraction stage

— Stage 2: MFCC Feature Extraction.

The detected speech segment undergoes frame-based processing with 128-sample Hamming windows and 50% overlap. For each frame, a 128-point FFT is computed, followed by mel-filterbank application (12 filters, 300-3800 Hz range) and DCT to obtain 8 MFCC coefficients. The adaptive binning aggregation method then converts the variable-length MFCC matrix into a fixed 64-dimensional feature vector.

— Stage 3: LogNNet Classification.

The feature vector is processed through the LogNNet 64:33:9:4 architecture, consisting of a 33-row reservoir layer with chaotic mapping weights, followed by a 9-neuron hidden layer and 4-output softmax layer for command classification.

Efficient utilization of limited RAM is a critical factor for embedded systems. Table 4 presents a comparison of memory requirements for different MFCC feature aggregation methods.

Table 4. Memory requirements for different MFCC aggregation methods.

| Aggregation method | Feature Vector Size | Additional RAM for Computation | Total RAM Usage |
|---|---|---|---|
| Basic statistical features | 32 elements × 4 bytes (type float) = 128 bytes | 40 bytes local variables | 168 bytes |
| Temporal dynamics | 48 elements × 4 bytes (type float) = 192 bytes | 52 bytes local variables and 388 bytes temporary buffers – 440 bytes | 632 bytes |
| Windowed statistics | 128 elements × 4 bytes (type float) = 512 bytes | 44 bytes local variables | 556 bytes |
| Adaptive binning | 64 elements × 4 bytes (type float) = 256 bytes | 20 bytes local variables | 276 bytes |

\* Excluding input MFCC data (4000 bytes), which is common to all methods.

Comparison of aggregation methods shows that the Basic Statistical Features method is the most compact in terms of output vector and the least in terms of RAM (168 bytes), but loses temporal information; the Temporal Dynamics method partially preserves it, but due to additional buffers for derivatives and local variables, it requires a total of 632 bytes, which significantly loads the memory. Windowed Statistics forms the largest feature vector with a total RAM cost of 556 bytes, while the Adaptive Binning method provides the best balance - a moderate vector size, with minimal overhead costs and a total RAM consumption of 276 bytes.

Taking into account the 32 KB RAM limitation and the need to reserve memory for other system components (8 KB audio buffer, 2 KB FFT buffers, etc.), the Adaptive Binning method is the optimal solution.

Table 5. RAM memory allocation by component.

| Component | RAM (bytes) | Description |
|---|---|---|
| **Audio Processing** | | |
| Circular buffer | 8000 | 4000 samples × 2 bytes (int16_t). Ring buffer for continuous audio capture at 8 kHz sampling rate, stores 0.5 seconds of audio |
| VAD state variables | 88 | Finite state machine (4 states), counters etc. |
| Energy history | 200 | 50 frames × 4 bytes (float). Sliding window for adaptive threshold calculation |
| **MFCC Processing** | | |
| FFT buffers | 2048 | 128 complex samples × 2 × 8 bytes (double) |
| Mel filterbank | 3072 | 12 filters × 64 frequency bins (half of 128-point FFT) × 4 bytes (float) |
| DCT matrix | 384 | 8 output coefficients × 12 input mel-energies × 4 bytes (float) |
| MFCC frames (dynamic) | 4000 * | Maximum 125 frames × 8 coefficients × 4 bytes (float). Stores complete MFCC matrix for one command |
| Feature vector | 256 | 64 features × 4 bytes (float). Final aggregated vector using adaptive binning |
| **LogNNet Model** | | |
| LCG parameters [16] | 8 | 4 × 2 bytes (int). Linear Congruential Generator constants |
| Reservoir normalization | 396 | 33 reservoir neurons × 3 statistics (max, min, mean) × 4 bytes (float). Pre-computed from training data |
| MLP weights (33→9) | 1188 | 33 hidden neurons × 9 hidden neurons × 4 bytes (float) |
| MLP weights (9→4) | 144 | 9 hidden × 4 output neurons × 4 bytes (float) |
| Runtime buffers | ~ 300 | Sh[34], Sh2[10], Sout[4] × 4 bytes and local variables |
| **ArduinoFFT Library** | ~500 | FFT object and methods |
| **System Variables** | ~ 1000 | Global state, interrupt handlers, stack allocation, Arduino core variables |
| **Total Measured** | 18016 ** | Arduino IDE compiler output (actual runtime allocation) |
| **Utilization** | 54.9% | |

\* Dynamically allocated only when processing a command.
\*\* Total is less than sum due to dynamic allocation of MFCC frames.

The implementation achieves memory utilization at 54.9% of available RAM. This provides sufficient headroom for wireless communication tasks when utilizing the WiFi module.

The observed consumption of about 50 KB of Flash is explained by the fact that the firmware includes not only application logic, but also a significant infrastructure layer. The assembly contains framework components and hardware-dependent drivers (UART, ADC, timers), startup code and elements of the C/C++ runtime environment. The ARM Cortex-M0+ architecture does not have a hardware coprocessor for floating-point operations, so float calculations are performed in software; together with mathematical libraries, including functions used in calculating MFCC, this forms a significant share of the final volume. An additional contribution is made by the FFT implementation, interrupt handlers, audio stream buffers, VAD logic and model data structures located in program memory. These factors naturally increase the size of the binary file.

## 4  Conclusion

This work presents a co-designed approach to speech command recognition for severely resource-constrained embedded systems, demonstrating that reservoir computing principles can effectively compete with deep learning methods when memory is limited. The key contributions include: systematic evaluation of MFCC aggregation methods revealing adaptive binning as optimal for embedded deployment, empirical proof that LogNNet achieves 92% speaker-independent accuracy with 10× fewer parameters than conventional DNNs, and complete implementation fitting within 18 KB RAM on a Cortex-M0+ processor without DSP support.

The results validate reservoir computing as a viable alternative for edge AI applications where traditional neural networks are impractical. Future work should explore dynamic vocabulary loading, investigate performance under extreme acoustic conditions, and extend the approach to continuous keyword spotting. The open-source implementation provides a foundation for deploying intelligent voice interfaces across the spectrum of IoT devices, from smart sensors to wearable electronics.

## 5     Discussion.

The implemented system demonstrates several practical advantages for real-world deployment. First, the 54.9% RAM utilization leaves sufficient headroom for wireless communication protocols, enabling integration with the NINA-W102 module's Wi-Fi and Bluetooth capabilities without memory conflicts. Second, the dual-mode operation (idle monitoring vs. active processing) aligns well with battery-powered IoT constraints, as the system performs only lightweight VAD in standby mode, activating full processing pipeline only upon speech detection.

Compared to existing embedded speech recognition solutions, our approach offers distinct trade-offs. CNN-based systems demonstrate high accuracy on more powerful platforms: DS-CNN on STM32F746 achieves 94.4% accuracy but requires 80KB RAM [17], while attention-based CNNs on STM32F469 reach 93.9% with 128KB RAM requirements [18]. Our LogNNet implementation achieves comparable 90% on-device accuracy with only 18KB RAM (3-5× reduction) and no DSP requirements, making it suitable for cost-sensitive applications where Cortex-M0+ processors are preferred over Cortex-M4F variants.

Thus, the implementation on the Arduino platform demonstrated the practical applicability of the proposed algorithm for wireless embedded systems with strict constraints on memory, computational power, and energy consumption. The results confirm the effectiveness of the LogNNet architecture and adaptive MFCC feature binning method for creating autonomous voice command recognition nodes in wireless personal communication systems and IoT networks.

## Acknowledgments

Special thanks to the editors of the journal and to the anonymous reviewers for their constructive criticism and improvement suggestions.

This research was supported by the Russian Science Foundation (grant no. 22-11-00055-P, https://rscf.ru/en/project/22-11-00055/, accessed on 10 June 2025).